# A call for scientists to halt the spoiling of the night sky with artificial light and satellites

Fabio Falchi[1,2,*], Salvador Bará[3], Pierantonio Cinzano[2], Raul C. Lima[4,5], Martin Pawley[6]

[1]Departamento de Física Aplicada, Universidade de Santiago de Compostela; Santiago de Compostela, Galicia, Spain

[2]ISTIL Istituto di Scienza e Tecnologia dell'Inquinamento Luminoso – Light Pollution Science and Technology Institute; Thiene, Italy

[3]Independent Scholar; Santiago de Compostela, Galicia, Spain

[4]Física, Escola Superior de Saúde, Politécnico do Porto, Portugal

[5]IA – Instituto de Astrofísica e Ciências do Espaço, Univ Coimbra, Portugal

[6]Agrupación Astronómica Coruñesa Ío, 15005 A Coruña, Galicia

*Corresponding author. Email:  falchi@istil.it

*Unfettered access to dark night skies is rapidly diminishing, due to light pollution and satellite mega-constellations tracks. Scientists should wake up and do more to stand up to Big Light and Big Space and preserve this natural resource.*

Light pollution, produced both at Earth surface and from Low Earth orbit (LEO) satellites, is rapidly increasing. In the case of light at night, light is considered by the general public as a positive thing *per se*, with no side effects, notwithstanding the evidence produced by scientific research in the last decades. Yet Artificial Light At Night (ALAN) is an anthropogenic pollutant, as defined since 1979 by the United Nations[1]:

> 'Air Pollution' means the introduction by man, directly or indirectly, of substances or *energy* into the air resulting in deleterious effects of such a nature as to endanger human health, harm living resources and ecosystems and material property and impair or interfere with amenities and other legitimate uses of the environment, and "air pollutants" shall be construed accordingly" (italics added) where[2] 'Energy' is understood to include heat, *light*, noise and radioactivity introduced and released into the atmosphere through human activities" (italics added).

The above definitions are perfectly compatible with the description of light pollution in terms of the volume concentration of anthropogenic photons in Earth's atmosphere[3,4]. The scientific literature on the negative consequences that light pollution has on human and animal behaviours and physiology is rapidly growing and does not allow ALAN to be considered as something other than a pollutant. Of course, sometimes pollution is an unavoidable by-product of a necessary thing. As an example, using a gasoline engine of an ambulance to save lives is an indisputable benefit, but this does not suppress the pollution produced while driving it. In this case, the advantage is overwhelmingly higher than the disadvantage of $NO_X$, $CO_2$, particulate matter, and other pollutants generated by the ambulance functioning. Also, it appears that some of the supposed benefits of lighting are questionable, for example, greatly enhancing public safety, reducing the justification for the lighting in the first place[5,6].

Astronomers are well aware of the negative consequences of ALAN, the increase of the background (really, the foreground) radiance that hinder, more or less depending on its intensity, the possibility to do astronomical research using the full potential of the telescopes. They know this well, as over the time they have progressively located new telescopes further and further away from cities. Today, due to the rise of light pollution, there are almost no more remote places available on Earth that meet simultaneously all the characteristics needed to install an observatory (namely, the absence of light pollution, a high number of clear nights, and good seeing). Most of the 3 metre and larger telescopes operate under night skies that surpass the IAU maximum allowable limit for acceptable interference by artificial light[7]. Given the paucity of sites, it should be expected that we all fight tooth and nail to preserve the night sky darkness of actual and potential future sites.
Unfortunately, reading the reports resulting from *Dark and Quiet Skies for Science and Society* conferences[8,9] and other reports on these subjects, we are very pessimistic about the path being followed by an important part of the scientific community (and other actors) that works and has responsibilities on these areas of research. Let's recall what has happened in other fields in the last decades, such as the findings related to the tobacco smoke (active and passive), acid rain, climate warming, diesel emissions, asbestos, ozone hole, silicosis, PFAS (forever chemicals), opioids, and sugar, to name only some. Every time that some health or environmental issue arises and starts to be addressed in the scientific literature, the 'machine of doubt' is put into action by the polluters to stop, or at least delay by years or decades, the adoption of countermeasures and rules to protect human health and the environment[10,11]. The strategy is always the same. The polluters argue that there is no evidence, or that the evidence is weak, or that causation is not proven that their product creates health or environmental problems. This also done by funding and producing deliberate contrary 'scientific' research and publications. The resulting procrastination of the adoption of limits on pollution has caused millions of deaths and has produced the over-polluted world we have today. Unaware scientists try to play a fair game against which Big Oil, Big Tobacco, Big Pharma, Big Sugar, and so on, simply work to increase profits, by skipping rules and trying to avoid their adoption.

Are all astronomers and other scientists working on light pollution issues (including light pollution from satellite constellations) aware of those episodes, and shouldn't we all keep our guard? Will Big Light act differently from their cousins? Big Light has already been proven to have protected their sales with the Phoebus cartel, made up by the main lamp manufacturers, which one century ago forced its members to reduce the lifespan of incandescent light bulbs from 2,500 hours to about 1,000[12]. The 1,000-hour life of incandescent bulbs is what we still have today. This was evidently made to keep the sales high, with a substitution market 2.5 times faster, and therefore the companies richer than otherwise would be the case. Now both indoor and outdoor lighting is transitioning fast toward LEDs, with predicted lifespan of 50,000–100,000 or more hours. This means that with a typical all-nighter, dusk-to-dusk lighting of 4,200 hours/year, a 100,000-hour LED lamp-post will last for about 24 years, compared to the about 6 years of High Pressure Sodium (HPS) lamps. Will Big Light settle for selling a quarter of the amount of lamps, or will they search to find evermore new ways to light up the night? We are experiencing more and more artificial light where, and when, there was little or none previously: two fixtures for each pole, one for the road, one for the walkway; road lighting outside settlements; the lighting of secondary roads and bridges; lighting of every monument and every building; ski slope lighting; lighting of waterfalls, high mountains, beaches, and others. Also, as a side consequence, even if LED is more efficient than previous technologies, the opportunity to use much less energy than before is also failing due to the higher amount of light being used. It surely should be mandatory for governments to take immediate actions to limit and lower the total amount of ALAN, similar to what they did to control other atmospheric pollutants. But trying to decrease light pollution or even just stop its growth using only prescriptions on fixtures or single installations is doomed to failure. This traditional and well-intentioned approach to halt light pollution has failed so far. In fact, light pollution has been increasing rapidly. Every added light, however low its pollution is for the desired task, whichever orientation towards the sky or ground it has, will increase the total amount of ALAN. Something we now know how to analyse and control with new strategies[13,14,15]. For example, it is possible to determine the contribution of each area that produces light pollution to the total light pollution (intended as, e.g. zenith brightness, average radiance over the entire sky hemisphere, average radiance in the first 10 degrees above the horizon, average radiance at 60 degrees zenith distance, horizontal irradiance) experienced in a location, be it a national park, a protected area, an observatory or any site we want to study.

What about Big Space? Regarding the impacts of Low Earth orbit (LEO) satellites on the night sky and science, it is similarly naive to hope that the skyrocketing space economy will limit itself, if not forced to do so, to counter the new environmental and security issues raised by the new mega-constellations of satellites. The approach of part of the scientific community to this rising problem is to mitigate the impact of the light reflected toward the night hemisphere by these satellites, by lowering their brightness, by closing the shutters of telescopes' instruments when they are in the field of view, by pointing telescopes where there are no satellites, trying to skip them. This might mitigate some of the problems, but will not solve them, nor solve all the related problems. The loss of the natural aspect of a pristine night sky for all the world, even on the summit of K2 or on the shore of Lake Titicaca or on Easter Island is an unprecedented global threat to Nature and Cultural heritage. If matters continue, dozens, hundreds, thousands of satellites will be seen crossing the skies at a given moment and no human being will be able to admire the night sky as it was always possible to do. If not stopped, this craziness will become worse and worse. Even if the main actors in this field will be able to dim the brightness of its satellites far beyond the naked-eye limit, many other problems will remain unresolved, including orbital traffic concerns, atmospheric pollution from debris and from rocket exhaust gases, and, of course, the increase of sky brightness background[16] (that even in this case is more precisely a foreground). The exploitation of the low Earth orbit mega-constellations includes super-fast speculative financial transactions and battlefield management. Such an unprecedented escalation should be stopped at the outset and regulated, with some suggestions given below. Today, contrary to what has allowed for decades to grow in light pollution on the Earth's surface, we cannot speak of ignorance of the impacts, losses and risks of mega-constellations. However, we are permitting this escalation. By engaging in dialogue with companies, instead of dialogue with States (or demanding from them) or with international regulating entities, we are allowing the interested party to self-regulate, replacing the role of a regulatory State, which must be the guarantee of the well-being of societies. Mitigation is not regulation. Regulating may include the exclusion of a practice for the benefit of societies. No company will do this deliberately. On the scales of immediate or long-term benefits and harm to society, and despite the popularity of satellite mega-constellations, we must not reject the possibility of being banned. On the contrary, we believe that the impacts and risks are too high for this possibility to be ruled out.

Whatever the particular situation in each country there are a few key actions that astronomers worldwide and their academic and professional societies may adopt: (i) explicitly reaffirming the essential value of the unpolluted night sky as a common asset of humankind, (ii) actively promoting the preservation and — wherever necessary — the recuperation of the starry skies as a first-class strategic goal, not secondary nor ancillary to any for-profit exploitation of natural resources, including the LEO space zone, (iii) realize that there is a very likely basic and unavoidable conflict of goals between scientific and industrial activities that cannot be satisfactorily solved by a naïve 'all-stakeholders' approach, and (iv) recall that these problems are socio-political, not technological in nature, and to act in consequence. Decided action should be taken in all countries, more urgently so in those who bear a larger share of responsibility in the present process of deterioration of the global night sky. Immediate steps may include reinforcing public appeals for tightening the criteria for authorizing massive satellite launches (e.g. US FCC-FAA ones), subjecting this kind of activity to rigorous environmental impact assessment, and repealing regulations that hinder preserving our global commons (such as former US administration Executive Order 13914 of April 6, 2020, still in force[17]).

A cap strategy should be enforced, a decades-old way successfully used to control most pollutants, both for artificial light production and orbiting satellites. Approaching these caps, actions should be taken to re-enter in them (for instance, by diminishing the light produced elsewhere, if some new installation is to be built; deorbiting old satellites and space debris to allow for new ones). If the reasonable caps are already surpassed, as it seems in both light pollution and satellite pollution, actions should be taken to remediate.

All relevant actors should be called to rebuilding international cooperation and agreement in order to avoid escalation. Scientists and scientific societies are entitled to actively promote this stance before their governments and regulatory organizations. In the international arena the scientific community, with long-standing and strong personal and professional collaborative ties extending worldwide, is in an optimum position to counteract the present trend towards unilateralism and conflict. As it is not too late to stop this, we as scientists and first as citizens should act in the first person to stop this attack, from above with satellites and from below with ALAN, on the natural night and on the intangible cultural heritage of humankind's starry skies[18]. Now is time to consider the prohibition of mega-constellations and to promote a significant reduction in ALAN and the consequent light pollution. Our world definitely needs a New Deal for the night.